# Significant enhancement of hole mobility in [110] silicon nanowires compared to electrons and bulk silicon

A. K. Buin,[*] A. Verma,[†] A. Svizhenko,[#] M. P Anantram[*,1]

*Nanotechnology Program, University of Waterloo, 200 University Avenue West, Ontario N2T 1P8, Canada

[†] Dept of Electrical Engineering, Texas A&M University-Kingsville, TX, USA

[#] Silvaco Data Systems Inc., 4701 Patrick Henry Drive, Santa Clara, CA, USA

[1] anant@uwaterloo.ca

**Abstract:**

Utilizing sp3d5s* tight-binding band structure and wave functions for electrons and holes we show that acoustic phonon limited hole mobility in [110] grown silicon nanowires (SiNWs) is greater than electron mobility. The room temperature acoustically limited hole mobility for the SiNWs considered can be as high as 2500 $cm^2$/Vs, which is nearly three times larger than the bulk acoustically limited silicon hole mobility. It is also shown that the electron and hole mobility for [110] grown SiNWs exceed that of similar diameter [100] SiNWs, with nearly two orders of magnitude difference for hole mobility. Since small diameter SiNWs have been seen to grow primarily along [110] direction, results strongly suggest that these SiNWs may be useful in future electronics. Our results are also relevant to recent experiments measuring SiNW mobility.

With the semiconductor industry fabricating devices with feature sizes in the tens of nanometers, there is a potential for silicon nanowire (SiNW) devices to play an important role in future electronics, sensors, and photovoltaic applications. Since the band structure of SiNWs varies vastly with their physical structures,[1-4] it provides a tantalizing possibility of utilizing different SiNWs within the same application to achieve optimum performance. For example, a difference in the physical structure of the SiNWs results in a difference in sensing properties.[5] This implies that in order to bring the promise of SiNWs to fruition, it is necessary to characterize a vast number of SiNWs with different cross-section shapes and sizes, and axis orientations, for their electronic properties. In particular, one of the most important properties is the low-field mobility. An accurate knowledge of the low-field mobility is important in order to select the right SiNW for a particular application. The low-field mobility is strongly influenced by acoustic phonons,[6] surface roughness scattering,[7] and impurity scattering.[8] Surface roughness and impurity scattering are to large extent controllable parameters. Phonon scattering on the other hand is intrinsic, and hence places an upper bound on the expected carrier mobility. Recent results suggest that electron mobility in SiNWs can be significantly increased by coating them with acoustically hardened materials,[8] largely by clamping the boundary, or by inducing a strain.[9] This in turn also makes a strong case for investigating the performance limitations of freestanding SiNWs.

Utilizing various boundary conditions, many researchers have computed the electron-phonon scattering rates in nanowires based on an effective mass equation.[10-16] Reference 17 is an exception that considered a tight-binding Hamiltonian for the electron, albeit with bulk-like confined phonons, where bulk quantized transverse phonon vectors emulate confined phonon behavior. Further, the analysis has been limited to consideration of the lowest conduction subband. It is without doubt that these methods make it easier to compute scattering rates. However, they come with a possibility of a compromise in accuracy, in particular, for holes where the valence subbands are closely spaced together and intersubband scattering may be significant. Accounting for intersubband hole scattering is one reason that makes it challenging to theoretically investigate hole transport in these dimensionally reduced structures, a task all the more imperative because of recently reported very high hole mobility[18] in experiments.

It is clear, therefore, that in order to reproduce an accurate description of the physical effects taking place in these confined structures and evaluate charge transport properties, it is important to take into account both charge and phonon confinement. Concomitantly, it is also important to investigate how these properties compare to bulk-phonon values with a more detailed treatment of these three-dimensional phonons. Bulk phonons not only provide an ease of modeling, but also make it easier to incorporate intersubband carrier scattering, which, as our results show, is very important even in many small diameter SiNWs. More importantly, consideration of bulk phonons address the scenario of a SiNW encapsulated within an acoustically similar material.

In this Letter we present results on a detailed computation of electron and hole low-field mobility for [110] axially oriented free standing SiNWs (henceforth referred to as [110] SiNWs in this paper) with diameters up to 3.1 nm and at various temperatures, where the principal charge scattering mechanism is through acoustic phonons. Both confined and bulk phonons are considered. The band structure for these SiNWs is determined by using a $sp^3d^5s^*$ tight-binding (TB) scheme and the confined acoustic phonon dispersion for each SiNW are obtained by solving the elastic continuum wave equation. Bulk phonon dispersion is assumed to be linear and a Debye cut-off energy is used to define the domain of bulk phonon wave vectors. Electron and hole - acoustic phonon momentum relaxation rates are calculated from the first order perturbation theory and deformation potential scattering. In computing the charge momentum relaxation rates for a SiNW from confined phonons, all acoustic phonon modes up to 70 meV[19-21] are taken into account, and TB electron and hole wavefunctions are incorporated. Finally, low-field mobility values are determined through momentum relaxation time approximation, and verified for electron-confined phonon interaction through ensemble Monte Carlo (EMC) simulations.

Fig. 1 shows a typical wire cross-section for the [110] SiNW. The periodicity of the SiNW along the lattice is given by the lattice constant $a/\sqrt{2}$, where $a$ is the lattice constant for bulk Si. To eliminate unphysical surface states, H atoms are used to passivate the Si dangling bonds at the SiNW edge. Within the $sp^3d^5s^*$ TB scheme, the Si-Si, and Si-H parameters are obtained from references 22 and 23, respectively. Fig. 2 shows the first few conduction and valence subbands for 2.4 nm diameter [110] SiNW. For the largest diameter SiNW considered in our work - 3.1 nm - the energy separation between the lowest two conduction subbands is 18 meV, while it is 17 meV for the highest two valence subbands. As is expected, as the SiNW diameter increases, the bandgap and intersubband spacing decreases. However, for [110] SiNWs this does not occur in the simple manner given by an effective mass Hamiltonian.

The electron and hole wavefunctions within the TB scheme for the [110] SiNWs are given by

$$\psi_{\nu}(k_z,\mathbf{r}) = \frac{1}{\sqrt{N}}\sum_{n,m} c_{\nu,m}(k_z)e^{ik_z na/\sqrt{2}}\varphi_m(\mathbf{r}-(\boldsymbol{\tau}_m \pm \mathbf{e}_z n\frac{a}{\sqrt{2}})) \quad (1)$$

where $\nu$ is the subband index, $k_z$ is the electron or hole wave vector along the z-axis, $\mathbf{r}$ is the radius vector, $N$ is the number of unit cells, and $\boldsymbol{\tau}_m$ is the basis vectors within the unit cell n. $\varphi_m$ are orthonormal Slater-type (Löwdin)[24] atomic orbitals, $c_{\nu,m}$ are expansion coefficients whose values are obtained within the TB scheme,[25] and $\mathbf{e}_z$ is the unit vector along the SiNW axis.

Different types of confined phonon modes such as dilatational, torsional, and flexural can exist within a SiNW.[26] Due to symmetry considerations, only dilatational phonon modes may contribute to intrasubband electron-phonon scattering.[27,28] Dilatational phonon modes correspond to a mixed nature of axial-radial atomic vibrations. Dispersion relationship for a coupled axial-radial dilatational mode is obtained from the elastic wave equation, and given by the Pochhammer-Chree equation, [26]

$$\frac{2k_l}{r}(q^2+k_t^2)J_1(k_l r)J_1(k_t r)-(k_t^2-q^2)^2 J_0(k_l r)J_1(k_t r)-4q^2 k_l k_t J_1(k_l r)J_0(k_t r)=0. \quad (2)$$

where, $r$ is the radius of the phonon box and $J_0$ and $J_1$ are Bessel functions of the first kind. Also, $k_{t,l}^2=\omega^2/\mathrm{v}_{t,l}^2-q^2$, where v represents the bulk acoustic velocity with subscripts $t$ and $l$ standing for transverse and longitudinal, respectively. Equation (2) is solved numerically to obtain $\omega(q)$ by considering the SiNW embedded within an equivalent circular phonon box with radius $r$. Fig.3 depicts the confined acoustic phonon dispersion for the 2.4nm diameter [110] SiNW with the band structure given in Fig. 2

The electron and hole – acoustic phonon scattering rates are calculated from the Fermi's golden rule and deformation potential approximation[29] as:

$$W(k_z,k_z^{'},\mathbf{q})=\frac{2\pi}{\hbar}\left|\left\langle \psi_\nu(k_z^{'}),N_{|\mathbf{q}|}+\frac{1}{2}\pm\frac{1}{2}\right|H_{\mathbf{q},e-ph}\left|\psi_\mu(k_z),N_{|\mathbf{q}|}\right\rangle\right|^2 \delta(E_\nu(k_z^{'})-E_\mu(k_z)\pm\hbar\omega(|\mathbf{q}|))$$

(3)

where $k_z$ and $k_z^{'}$ are the initial and final crystal momentum respectively, $N_q$ is the phonon equilibrium Bose-Einstein occupation number, $\mu$ and $\nu$ are initial and final subbands, respectively, $\delta$ is the Delta Dirac energy conserving function., and $\boldsymbol{q}$ is the phonon wave vector. $H_{\mathrm{q},e-ph}$ is the electron-phonon interaction Hamiltonian. One therefore obtains the momentum relaxation rates from confined acoustic phonons as:

$$W_{\mu,\nu}(k_z)=\sum_{k_z^{'}}W_{\mu,\nu}(k_z^{'},k_z)(1-\frac{k_z^{'}}{k_z})=-\frac{E_a^2}{4\pi\hbar^3\rho \mathrm{v}_s^4}\sum_n\sum_p\frac{q_p}{k_z}\gamma^{-1}\,[N(E_{ph}(q_p))+\frac{1}{2}\pm\frac{1}{2}]\,|S_{\mu,\nu}(q_p)|^2$$
$$E_{ph}^3(q_p)\ JDOS_\nu(k_z,q_p) \quad (4)$$

where $\mu,\nu$ indicate initial and final subbands, respectively. $JDOS_\nu(q,k,q_p)=\left|\partial(E_\nu(k_z\pm q_p)\pm\hbar\omega_n(q_p))/\partial q\right|^{-1}$ is the electron or hole - phonon joint density of states, $n$ indicates a summation over phonon modes, $p$ indicates summation over the roots of

the equation $\Delta E_{\mu,\nu}(k_z \pm q) \pm \hbar\omega_n(q) = 0$ (where $\Delta E_{\mu,\nu} = E_\nu - E_\mu$), and -/+ corresponds to absorption/emission. $E_a$ is the deformation potential, $\rho$ is the SiNW mass density, $\gamma$ is the normalization constant[30] and $k_l^2 = \omega^2 / \mathrm{v}_l^2 - q^2$. $S_{\mu,\nu}$, the overlap factor between subband $\mu$ and subband $\nu$ is given by

$$S_{\mu,\nu} = \sum_m c_{\mu,m}(k_z) c^*_{\nu,m}(k_z^{'}) J_0(k_l \rho_m) e^{iqz_m} \quad (5)$$

where $c_{\mu,m}$ and $c^*_{\nu,m}$ are defined through eqn.1.

In computing the momentum relaxation rates between electrons/holes and bulk acoustic phonons, the phonon dispersion is taken to be linear within the Debye approximation. The phonon dispersion is $\omega(|\mathbf{q}|) = \mathrm{v}_s |\mathbf{q}|$, where the longitudinal sound velocity $\mathrm{v}_s$=9.01*10$^3$ m/sec.[31] With the aid of the energy conserving Dirac Delta function and the discrete momentum conserving Delta function along the SiNW axis one obtains the charge momentum relaxation rates through bulk phonons, from initial subband $\mu$ to final subband $\nu$ as:

$$W_{\mu,\nu}(k_z) = \sum_{k_z^{'}} W_{\mu,\nu}(k_z^{'}, k_z)(1 - \frac{k_z^{'}}{k_z}) = -\frac{E_a^2}{4\pi\hbar^3 \rho \mathrm{v}_s^4} \int_\Omega \Delta E_{\mu,\nu}(k_z \pm q)^2 S_{1,\mu,\nu}(q_n, k_z \pm q)(N_{|\mathbf{q}|} + \frac{1}{2} \pm \frac{1}{2}) \frac{q}{k_z} dq \, . \quad (6)$$

In the above equation, $q_n = \sqrt{\left(\frac{\Delta E_{\mu,\nu}(k_z \pm q)}{\hbar \mathrm{v}_s}\right)^2 - q^2}$, and $S_{1,\mu,\nu} = \frac{1}{2\pi} \int_0^{2\pi} |S_{\mu,\nu}(\mathbf{q})|^2 d\varphi$ is the overlap factor integrated over phonon angular part in the confinement plane. The bulk Debye energy $E_D$ = 55meV[32] is utilized to define the domain of integration, $\Omega$. In general, $\Omega = \min, \max(1^{st}\,\mathrm{BZ}, \sqrt{q_n^2 + q^2} \le q_D)$, where $q_D$ is the Debye wave vector. In obtaining the above momentum relaxation rates prescription, we have neglected Umklapp processes.[11] Further simplifications allow us to obtain:

$$S_{1,\mu,\nu} = \sum_m |c_{\mu,m}(k_z)|^2 |c_{\nu,m}(k_z^{'})|^2 + \sum_{m,m^{'},m\neq m^{'}} A_{m,m^{'}} J_0(q_t \sqrt{\Delta x_{m,m^{'}}^2 + \Delta y_{m,m^{'}}^2}) e^{-iq\Delta z_{m,m^{'}}} \quad (7)$$

where $A_{m,m'} = c_{\mu,m}(k_z)c^*_{\nu,m}(k'_z)c^*_{\mu,m'}(k_z)c_{\nu,m'}(k'_z)$, and $\Delta j_{m,m'} = (j_m - j_{m'})$, where $j$ is x, y or z.

In evaluating the momentum relaxation rates using the above equations, we use a value of 9.5 eV for the electron deformation potential, and 5 eV for the hole deformation potential.[29] Electron and hole low-field mobility values are estimated based on relaxation time approximation with momentum relaxation time approximation (MRTA) and given by:

$$\mu = \frac{1}{\sum_i n_i}\sum_i \mu_i n_i, \quad (8)$$

where $n_i$ is the charge carrier population in subband $i$, and

$$\mu_i = \frac{2e}{k_B T m_{eff}} \frac{\int_{1^{st}BZ} E_i(k_z) W_i(k_z)^{-1} f_0(k_z) dk_z}{\int_{1^{st}BZ} f_0(k_z) dk_z} \quad (9)$$

for each subband $i$, where $W_i(k_z) = \sum_\nu W_{i,\nu}(k_z)$ and $f_0(k_z) \approx e^{-\frac{E_i}{k_B T}}$. $E_i$ is the electron/hole energy in subband *i,* measured with respect to the lowest/highest conduction/valence subband. Additionally, we verify low-field mobility values obtained using eqn. 9 for electron-confined phonons scattering by solving the Boltzmann Transport Equation using ensemble Monte Carlo (EMC) simulations.[29] EMC simulations utilize tabulated values of the electron band structure and scattering rates computed using results above. In performing EMC simulations, we consider SiNWs to be infinitely long and defect free. The temperature and electric field are assumed uniform. Given the nature of the phonons considered, only intrasubband scattering, restricted to the lowest conduction subband is considered, and electron energies are restricted to the bottom of the next higher subband. The bandstructure within that small region is divided into 12,000 grid points to minimize statistical noise at low electric fields.

The electron low-field acoustic confined phonon limited mobility computed using eqns. (4) and (8) are shown in Fig. 4. At room temperature, we find the mobility to scale approximately with diameter ($d$) and effective mass ($m_{eff}$) as:

$$\mu_e \propto d^2 m_{eff}^{-1.5} \qquad (10)$$

At low temperatures, we find that the scaling of mobility with effective mass remains the same, while the scaling with diameter is significantly weaker than $d^2$.

Bulk phonon limited electron low-field mobility is also shown in Fig. 4. We find that bulk phonons yield an electron mobility that is slightly larger than confined phonons. For the smallest diameter (1.27 nm) SiNW considered, the confined phonon mobility is about 38% lower than bulk phonon mobility, at room temperature. More significantly, this difference decreases to only 25% for the SiNW with a diameter of 3.1 nm. The diminished importance of phonon confinement at larger diameters is intuitively understood by noting that the phonon confinement is relatively weaker at 3.1 nm. As a result, confined phonons at 3.1 nm more closely resemble bulk phonons and therefore yield more comparable momentum relaxation rates when compared to the 1.27 nm diameter SiNW. We note that the bulk phonon scaling of mobility is qualitatively similar to that of confined phonons. The scaling of mobility with bulk phonons is consistent with the analytical results for scattering rates in reference 10. We remark that the electron mobility for the [110] SiNW is about four times larger than the mobility of [100] SiNW considered in reference 17. We also compare our electron mobility to reference 8, which found a mobility of around 336 $cm^2/Vs$ for a 4 nm diameter nanowire. This value is about 60% smaller than our mobility for the 3.1 nm diameter [110] SiNW. However, when scaled for the different diameters, effective masses, deformation potential and sound velocity, our results agree to within 15%.

The situation is somewhat different and more interesting for holes. Fig. 6 depicts the hole bulk acoustic phonon limited low-field mobility versus diameter for [110] SiNWs, at temperatures ranging from 77K to 300K. As with the case for electrons, hole mobility reduces as temperature increases because of an increase in momentum relaxation rates. Importantly, however, two things stand out. First, irrespective of the temperature, we find hole mobility to be significantly higher than electron mobility for a given SiNW. This is in contrast to bulk-Si where electron mobility is higher. This can also be seen in Fig. 7, which shows the variation of electron and hole mobility with temperature for an approximately 1.93 nm diameter [110] SiNW. Second, for the larger diameter SiNWs, we also find that hole mobility is higher than bulk-Si acoustic phonon limited hole mobility value by nearly a multiplicative factor of three[33] at 300K. These observed values of mobility are in line with experimentally observed results,[18] where peak hole mobility as high as 1350 $cm^2/Vs$ is observed and is attributed to reduced roughness scattering. Interestingly, by comparing results in Figs. 4 and 6 with reported bulk Si values at 77K, one can readily observe that hole mobility value (8486$cm^2/Vs$) is 55% lower than the bulk value (11481$cm^2/Vs$)[33] for the largest diameter SiNWs considered, while electron bulk-Si mobility value (23000$cm^2/Vs$)[34] is about eight times larger. Therefore, a reduction in temperature does not provide any mobility advantage over bulk-Si for these SiNWs.

Additionally at 300K, hole mobility values in Fig. 6 are nearly 2 orders of magnitude greater than the reported hole mobility for similar diameter [100] SiNWs.[17] This is primarily attributed to very heavy hole effective masses[35] in [100] SiNWs in contrast to [110] SiNWs. Fig. 8 depicts the hole effective mass for the top three valence subbands versus diameter for the [110] SiNW.

Unlike electrons for the SiNWs under consideration, developing a "rule-of-thumb" for the hole low-field mobility versus diameter is not straightforward. As can be seen in Fig. 6, this is due to the behavior of hole mobility for the ~1.93nm diameter SiNW. This behavior has two primary reasons. For the 1.93 nm diameter [110] SiNW, the highest two valence subbands are separated by an energy of approximately 15 meV, resulting in relatively high intersubband scattering. However, for the larger diameter SiNWs considered, these two subbands become nearly degenerate (energy separation is $\cong$ 1meV), thereby significantly reducing intersubband scattering due to phonon emission from the higher to the lower subband. Moreover, looking at Fig. 8, we also readily observe that the hole effective mass for the second subband decreases from approximately 0.25 for the 1.93 nm SiNW to 0.19 for the 2.4 nm SiNW. These factors result in a relatively sharp increase in hole low-field mobility for the larger diameter SiNWs compared to the 1.93 nm SiNW. Importantly, these results clearly demonstrate the importance of hole intersubband scattering because of the small energy differences between valence subbands. Therefore, while we have also calculated the hole mobility values due to the scattering arising from the dilatational modes of confined phonons, we do not report them because they grossly exaggerate the values of hole mobility since the dilatational modes are not sufficient to describe intersubband scattering.

In conclusion, we have evaluated electron and hole low-field mobility for [110] axially aligned SiNWs ranging in diameter from approximately 1 nm to 3.1 nm, utilizing bulk and confined acoustic phonons for electrons, and bulk acoustic phonons for holes, and including intersubband scattering. Hole acoustic phonon limited mobility for the 3.1nm diameter SiNW is found to be nearly 3 times the bulk-Si acoustically limited mobility value at room temperature. Additionally, hole mobility for the SiNWs considered in our work is also found to be higher than electron mobility, which is in contrast to bulk-Si, where electron mobility is higher. Moreover, both electron and hole mobility for the [110] SiNWs are greater than the reported [100] SiNW mobility values for similar diameters. Importantly, while electron and hole mobility values reported in this work help place an upper bound on the expected low-field mobility, they also clearly have technological implications for SiNW electronics for two important reasons. They point towards a preferential SiNW axis – [110], which is all the more important because small diameter SiNWs have been found to grow primarily along that direction.[36] They also demonstrate the possibility of high-speed devices where holes are the primary charge carriers.

**Figures:**

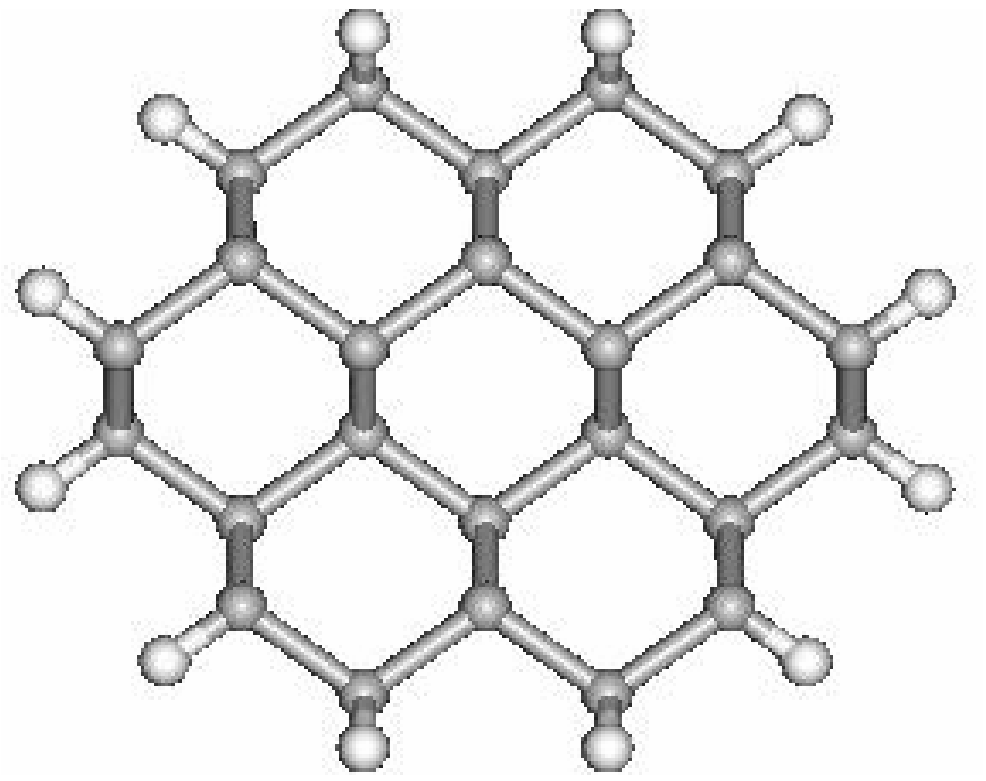

**Figure.1**

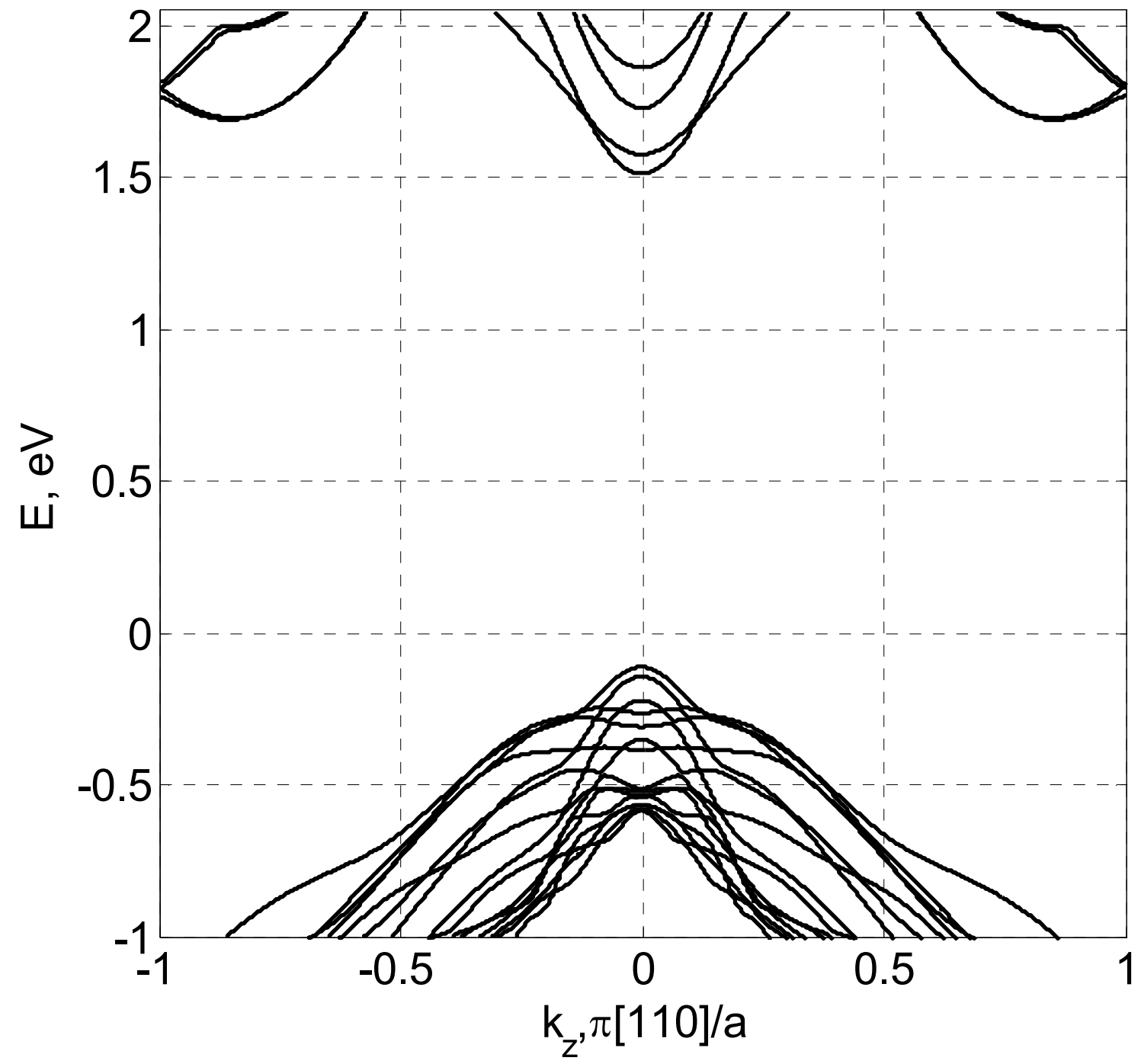

**Figure.2**

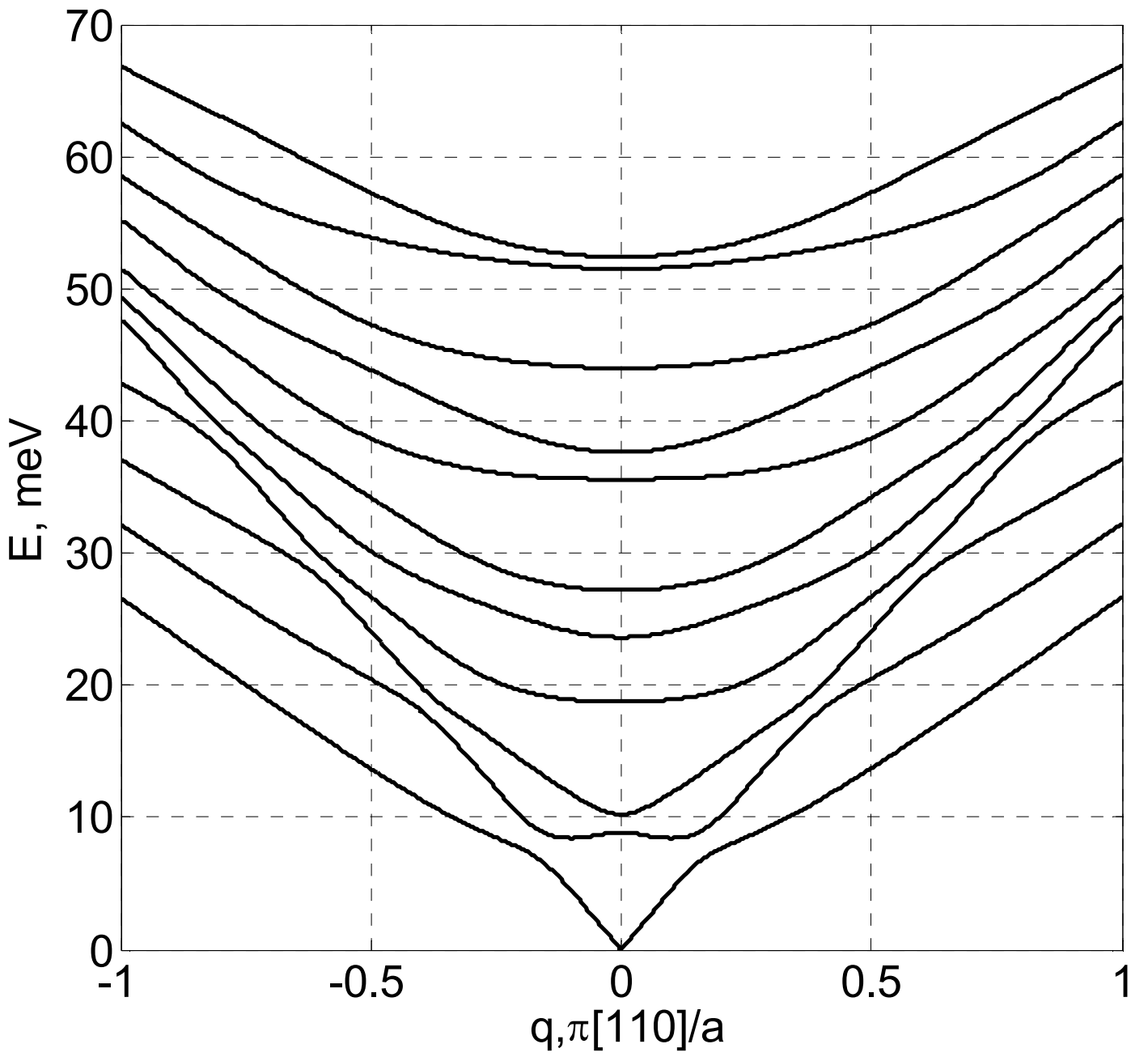

**Figure.3**

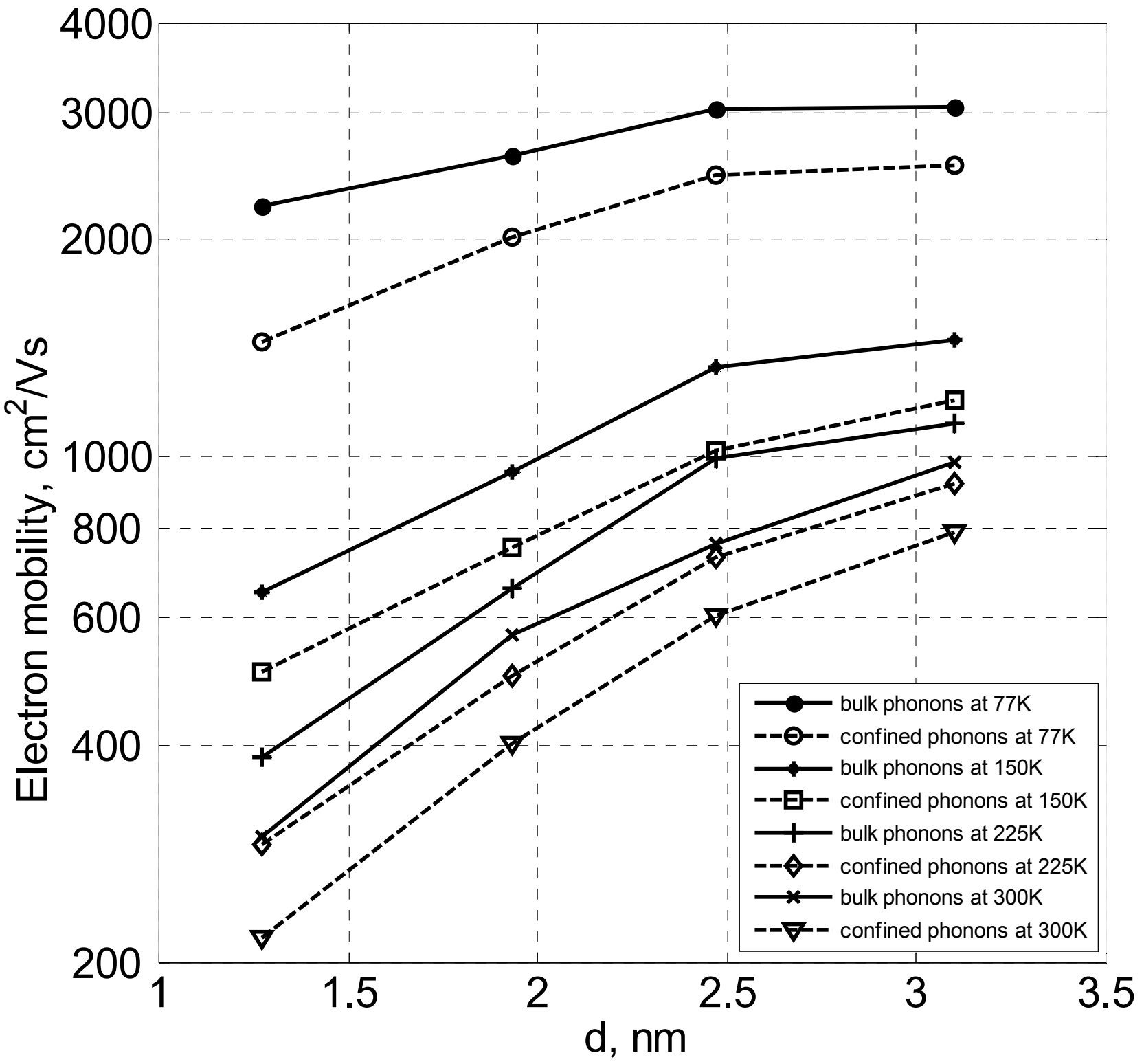

**Figure.4**

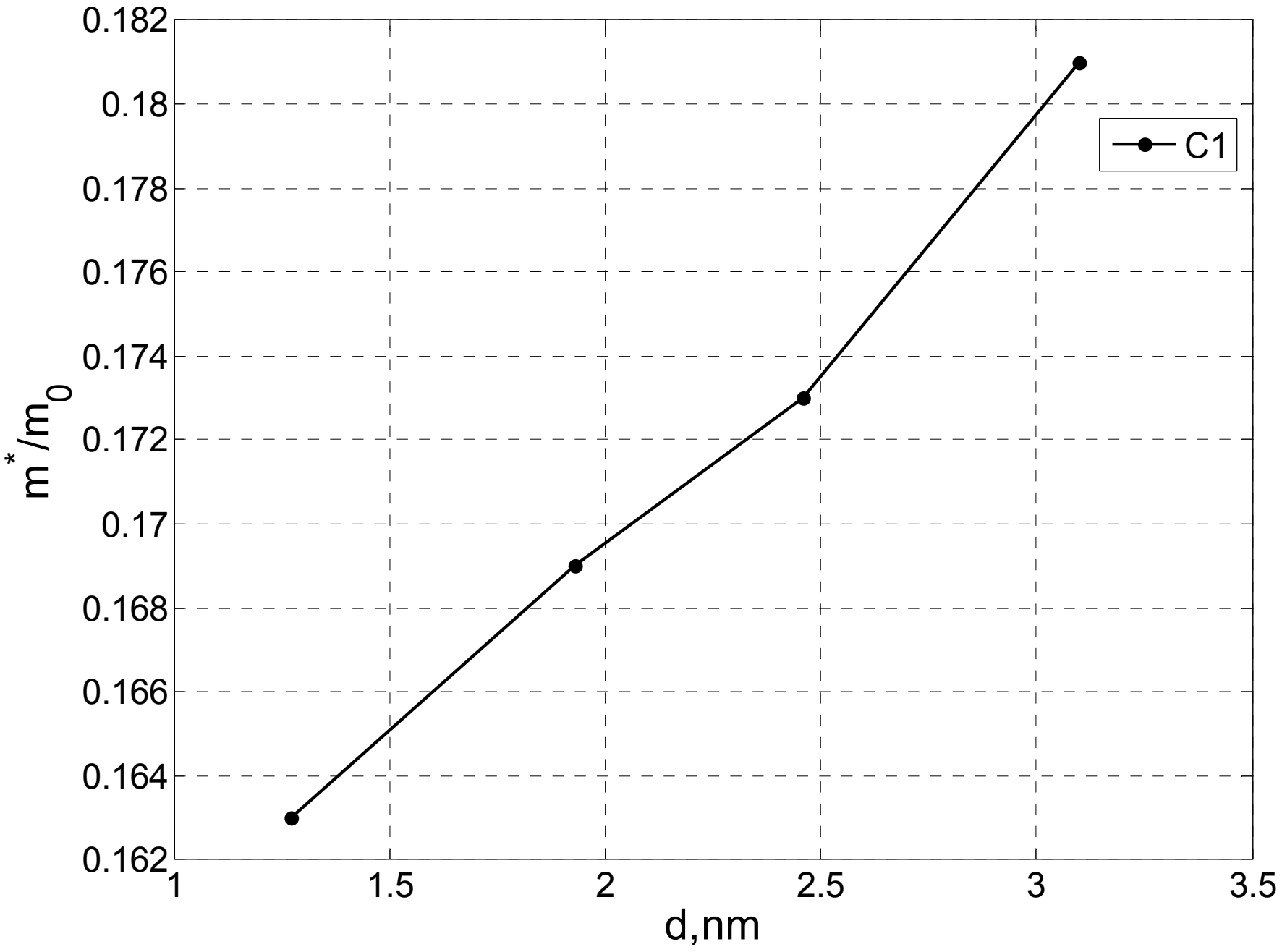

**Figure 5**

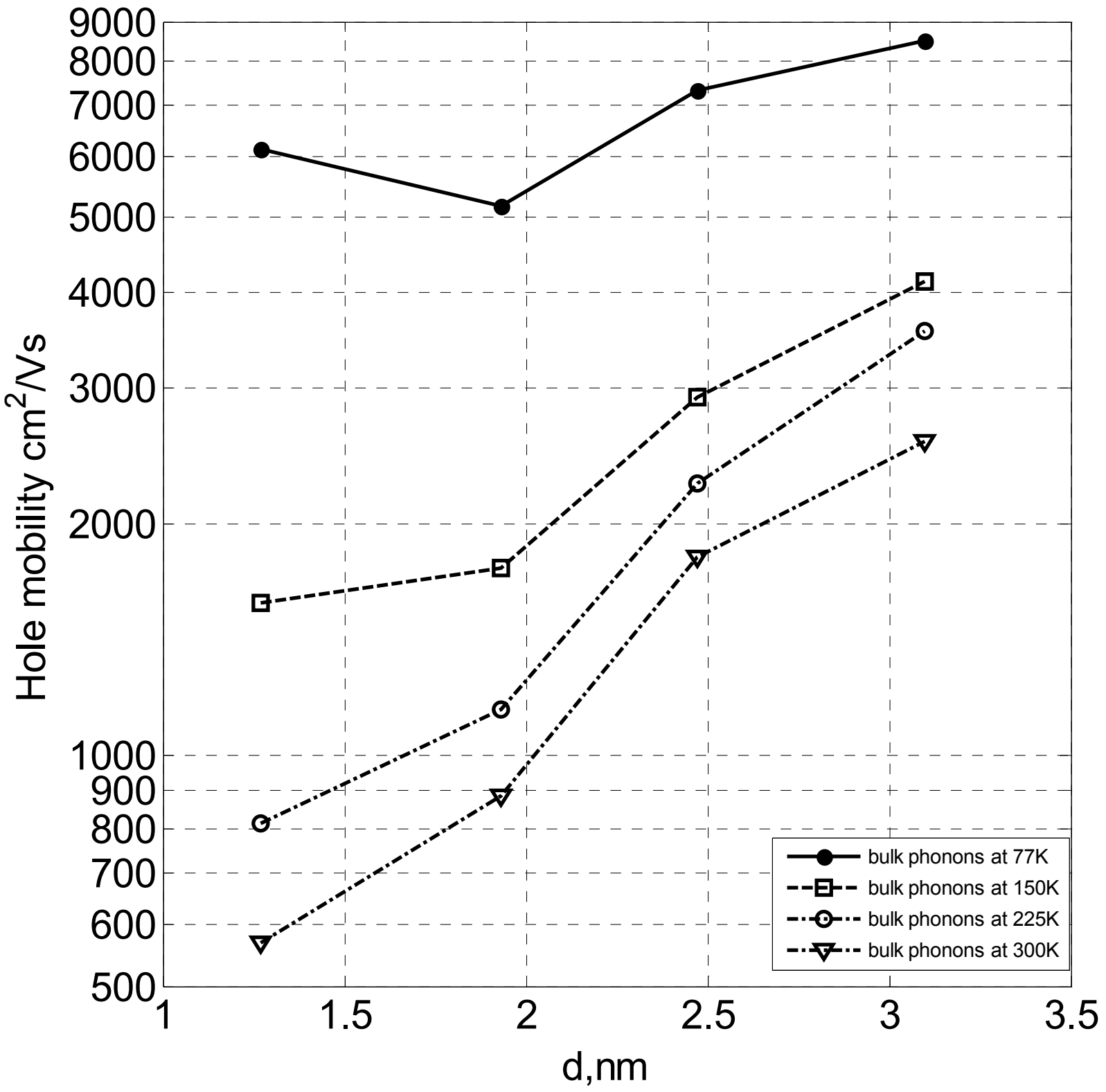

**Figure.6**

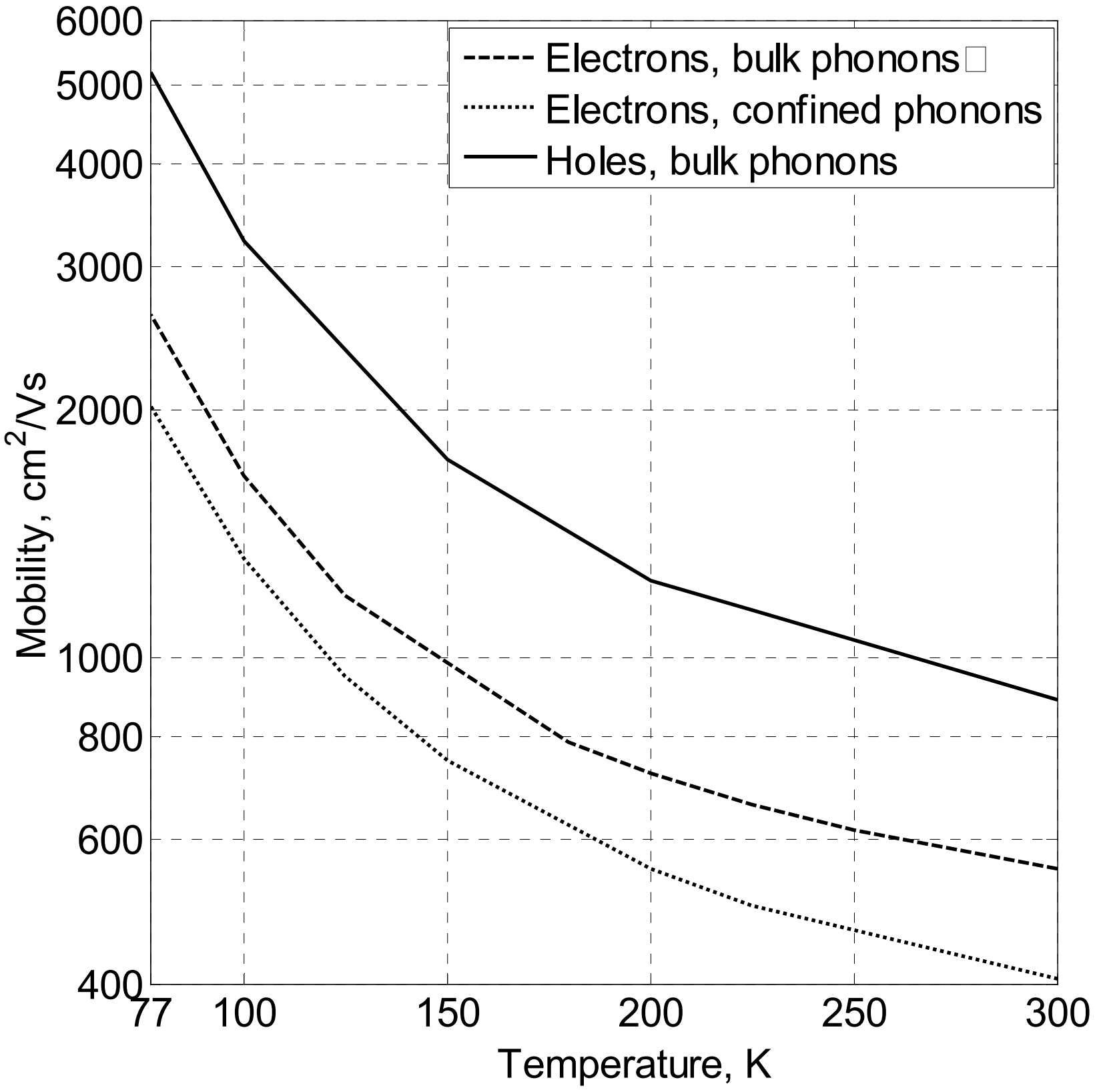

**Figure.7**

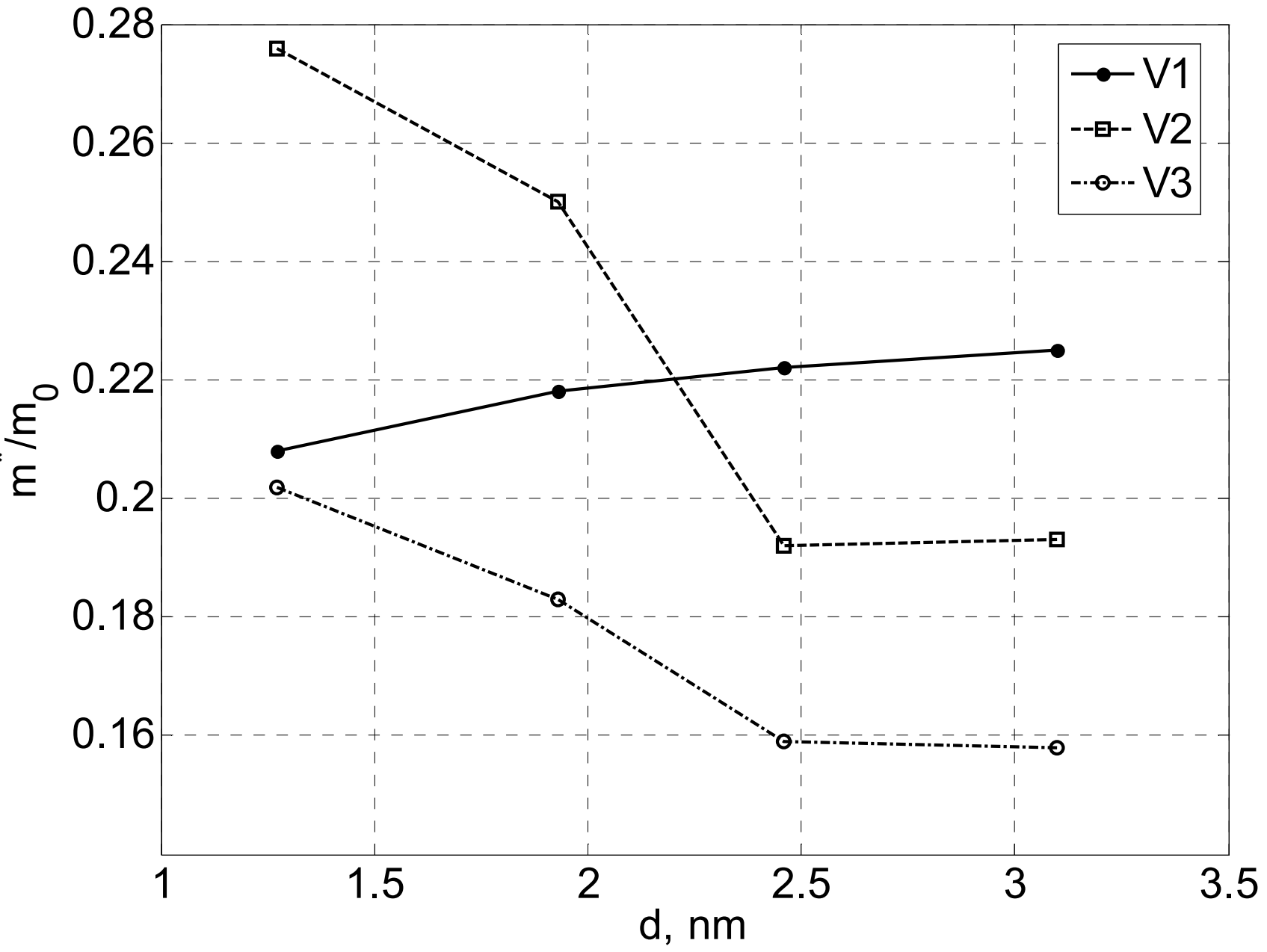

**Figure.8**

**Figure Captions:**
**Fig.1** Cross section of d=1.16 nm nanowire. Grey balls. Si atoms, white balls H atoms.
**Fig.2** Band structure of 2.4nm dimater [110] SiNW
**Fig.3** Dispersion curve for first 11 'dilatational' phonon modes for a 2.4 nm diameter [110] SiNW
**Fig.4** Electron Mobility(with bulk and confined phonons) versus [110] SiNW diameter for temperatures ranging from 77K to 300K
**Fig.5:** Electron effective mass versus [110] SiNW diameter. "C1" is first conduction band.
**Fig.6** Hole Mobility with bulk phonons versus versus [110] SiNW diameter for temperatures ranging from 77K to 300K
**Fig,7** Temperature dependence of mobility of 1.93nm diameter [110] SiNW
**Fig.8:** Hole effective mass versus [110] SiNW diameter. "V1" is the topmost valence band, "V2" is the next to the topmost valence band, "V3" is the lowest among 3.